# Distinguishing the importance of different charge trapping centers in CaF$_2$-based 2D material MOSFETs

Zhe Zhao[1,2], Tao Xiong [2], Jian Gong*[1,3] and Yue-Yang Liu*[2]

[1]School of Physical Science and Technology, Inner Mongolia University, Hohhot, China. E-mail: ndgong@imu.edu.cn
[2]State Key Laboratory of Superlattices and Microstructures, Institute of Semiconductors, Chinese Academy of Science, China. E-mail: yueyangliu@semi.ac.cn
[3]College of Physics and Electronic Information, Inner Mongolia Normal University, Hohhot, China. E-mail: gongjian@imnu.edu.cn

## ABSTRACT

Crystalline CaF$_2$ is drawing huge attentions due to its great potential of being the gate dielectric of two-dimensional (2D) material MOSFETs. It is deemed to be much superior than boron nitride and traditional SiO$_2$ because of its larger dielectric constant, wider band gap, and lower defect density. Nevertheless, the CaF2-based MOSFETs fabricated in experiment still present notable reliability issues, and the underlying reason remains unclear. Here we studied the various intrinsic defects and adsorbates in CaF$_2$/MoS$_2$ and CaF$_2$/MoSi$_2$N$_4$ interface systems to reveal the most active charge trapping centers in CaF$_2$-based 2D material MOSFETs. An elaborate Table that comparing the importance of different defects in both n-type and p-type device is provided. Most impressively, the oxygen molecules adsorbed at the interface or surface, which are inevitable in experiments, are as active as the intrinsic defects in channel materials, and they can even change the MoSi$_2$N$_4$ to p-type spontaneously. These results mean that it is necessary to develop high vacuum packaging process as well as preparing high-quality 2D materials for better device performance.

## 1. Introduction

Two-dimensional (2D) materials offer new possibilities for more Moore due to their ultra-thin thickness and smooth surface with no dangling bonds.[1–3] With the ultra-scaled channel, higher requirements are raised for the quality and reliability of gate dielectric materials.

To match the silicon technologies, oxides (such as SiO$_2$,[4] HfO$_2$,[5] and Al$_2$O$_3$[6]) are usually used, but these materials are non-layered, which makes it difficult to form a good interface with the 2D channels. To deal with the problem, 2D dielectrics such as h-BN have been studied.[7] However, the band gap (~6 eV) and dielectric constant (5.06 ε) of h-BN are not satisfying for dielectric materials.[8] Its band offset with 2D materials is not large enough, which will lead to many reliability problems.[9]

Excitingly, the recent experimental preparation of crystalline CaF$_2$ provides a strong support for the solution of this dilemma.[10,11] By using molecular beam epitaxy (MBE), crystalline CaF$_2$ can be grown on a silicon or germanium substrate.[12] It has a larger bandgap (12.1 eV) and larger dielectric constant (8.43 ε) than h-BN.[13] The grown CaF$_2$ is terminated by F atoms, which means that there is no dangling bond on its surface.[14] Another important point is that CaF$_2$ itself is stable in air, and is not easily dissolved in water.[15] CaF$_2$ can form good type I band alignment with many 2D materials, which means that it will be very advantageous as a gate dielectric of semiconductor devices.

Nevertheless, notable device reliability issues were still observed in CaF$_2$-based MOSFETs,[13,15–17] which contradicts the perfect electrical properties of CaF$_2$. For example, the $I_D$-$V_G$ hysteresis is significant (although lower than that in MoS$_2$/SiO$_2$ FET), and it shows obvious variability when the same device is operated at different scanning times. On the other hand, if different devices are operated under the same $V_D$, the $I_D$-$V_G$ characteristics such as on/off current ratio and subthreshold swing (SS) (150–90 mV dec$^{-1}$) differ greatly.[13] In addition, some devices with large negative threshold voltage ($V_{th}$) are prone to fail due to the bias overload of the CaF$_2$ layer. The physical origin of hysteresis and threshold voltage shift is widely attributed to the charge trapping and de-trapping of microscopic defects,[18–24] and the strength of the charge trapping effect is closely related to the type of defects.[25–28] Therefore, it is very urgent to distinguishing the activity of various defects in CaF$_2$-based transistors so that corresponding strategies can be proposed to deal with them.

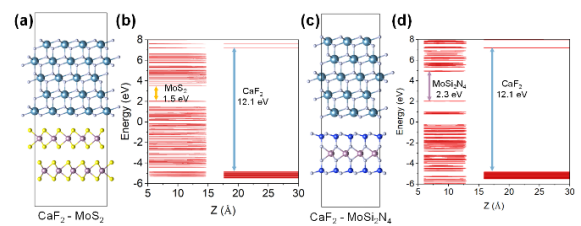

**Fig. 1** Atomic structure and type-I band alignment of the two kinds of interface models.

## 2. Method

Among the 2D materials, $MoS_2$ is one of the most widely used semiconductor. It has a direct band gap of 1.8 eV, and has been used to design high-performance electronic as well as optoelectronic devices.[29] On the other hand, there are also some new materials being synthetized, such as the $MoSi_2N_4$.[30] $MoSi_2N_4$ is very promising because of the excellent photocatalytic performance,[31] mechanical strength,[32] and electrical transportability.[33] Therefore, we construct both $MoS_2/CaF_2$ and $MoSi_2N_4/CaF_2$ interface models to make the simulation results representative. The lattice parameter of $CaF_2$, $MoS_2$ and $MoSi_2N_4$ is 3.90 Å, 3.16 Å, and 2.91 Å, respectively. To obtain good lattice matching, the primary cell of $MoS_2$ is repeated by five times to contact with the $CaF_2$ cell that is repeated by four times. The final $CaF_2$ deformation is only 1.28%. Similarly, the primary cell of $MoSi_2N_4$ is repeated by four times to contact the $CaF_2$ that is repeated by three times, and the $CaF_2$ deformation is only 0.52%. All the first-principles calculations are performed by the software PWmat.[34,35] The SG15 pseudopotential[36] is adopted, and the plane wave cutoff energy is 50 Ry. The Heyd–Scuseria–Ernzerhof (HSE)[37] functional is used in the calculation of electronic structures to improve the accuracy of calculations. The vdW force between the layers of the material is also considered.

## 3. RESULTS AND DISCUSSION

The interface models are shown in Fig. 1(a) and (c). A 5-layer $CaF_2$ is adopted because the experimental MBE grown $CaF_2$ is about 2 nm thick. The band alignments that manifested by the projected density of states (PDOS) are shown in Fig. 1(b) and (d). It can be seen that the VBM (valence band maximum) and CBM (conduction band minimum) are provided by $MoS_2$ and $MoSi_2N_4$, and the band offsets are greater than 2 eV, which makes charge tunneling difficult. This confirms that using $CaF_2$ as the gate of 2D material MOSFET is likely to obtain good device reliability.[38] Therefore, when considering practical applications, we believe that the reliability issues should stem from some intrinsic or external charge trapping centers.

Intuitively, we should first study the F vacancy defect in the $CaF_2$ layer. However, it has been demonstrated in experiment that $CaF_2$ is not easy to generate defects.[13] Besides, it has been proved by first principle calculation that even though F vacancies ($V_F$) and Ca vacancies ($V_{Ga}$) exist, there is no defect state near the band edge of channel material due to the large band offset between the two materials.[39] Consequently, we turn our attention to the trapping centers inside the channel material, in the semiconductor/dielectric interface, and at the dielectric surface. For $MoS_2$, we considered S vacancy defect ($V_S$), Mo vacancy defects ($V_{Mo}$), $MoS_3$ vacancy defect ($V_{MoS3}$) and $MoS_6$ vacancy defect ($V_{MoS6}$) at different spatial locations. On the other hand, considering that gas adsorption is very easy to occur in the process of device manufacturing, we also studied the water and oxygen molecules that adsorbed at different positions. For a more intuitive display of defects and adsorption, the related structural diagrams are shown in Fig. 2.

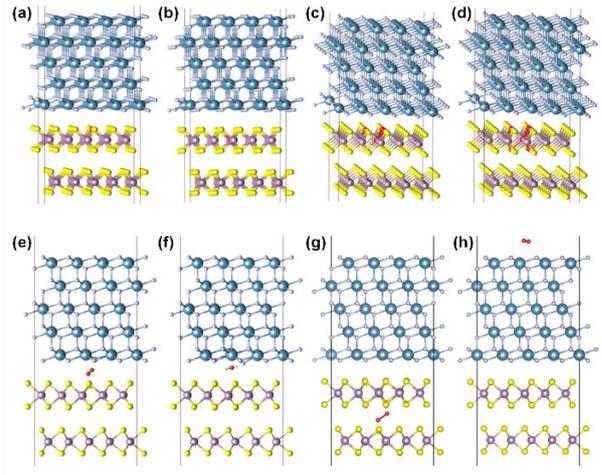

**Fig. 2** The different defects in $MoS_2$ from (a) to (d) are $V_S$, $V_{Mo}$, $V_{MoS3}$, and $V_{MoS6}$ defects. Oxygen molecules(e) and water molecules(f) are adsorbed between the $CaF_2$-$MoS_2$ interface, respectively. Oxygen is adsorbed on the interlayer of $MoS_2$(g) and the surface of $CaF_2$(h), respectively. The atoms highlighted in red in the figure represent defects and adsorption sites.

The energy level distribution of different defects in $MoS_2$ is shown in Fig. 3. First, there is an occupied defect state denoted by d1 for the $V_S$ in $MoS_2$, whose energy is 0.38 eV below VBM, and there are two empty defect states with similar energy denoted by d2, whose energy is 0.57 eV below CBM. According to charge transfer theories, the charge trapping rate will decrease exponentially with the increasing energy barrier between the initial and final electronic states, thus we can consider that only the defect levels that locate less than 1 eV away from the Si band edge are active trapping centers. Therefore, it can be concluded that d1 is an important hole trapping state when negative gate voltage is applied, and d2 is an important electron trapping states when positive gate voltage is applied. Similarly, the Mo vacancy are active in trapping holes and electrons, but they not as active as the S vacancy in electron trapping, because the $V_{Mo}$ defect levels are farther away the CBM. In addition to the common $V_S$ and $V_{Mo}$, experiments have reported that complex vacancy defects (such as V$_{MoS3}$ and V$_{MoS6}$) are found in MoS$_2$.[40] These two complex vacancies contain many dangling bonds, and thus can introduce a series of defect states (up to 13) that locate either close to VBM or to CBM. Consequently, they will be very active charge trapping centers. Nevertheless, the formation energy of these complex defects is very high, which makes them low in density. More details of the defect levels have been listed in Table 1.

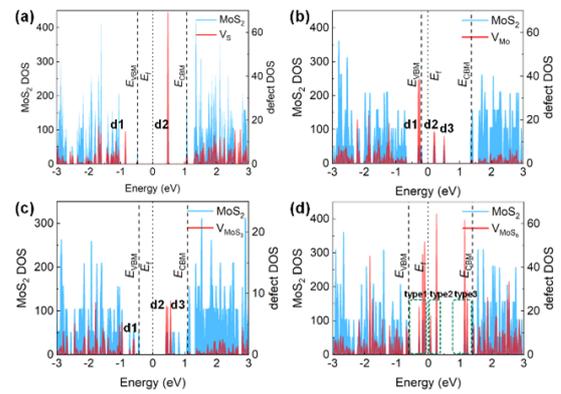

**Fig. 3** The energy level distribution of different defects. (a) S vacancy ($V_S$), (b) Mo vacancy ($V_{Mo}$), (c) $MoS_3$ vacancy ($V_{MoS3}$), and (d) $MoS_6$ vacancy ($V_{MoS6}$).

**Tab. 1** Importance of different trapping centers in CaF$_2$-MoS$_2$.

| Defect Types | Defect State | ΔE-VBM (eV) | ΔE-CBM (eV) | NBTI Importance | PBTI Importance | Formation/ Adsorption energy (eV) | Comprehensive Importance |
|---|---|---|---|---|---|---|---|
| V$_S$ | d1 | -0.38 | -1.91 | ✓ | ✗ | 2.91 | ✓ |
|  | d2 | 0.95 | -0.57 | ✓ | ✓ |  |  |
| V$_{Mo}$ | d1 | -0.06 | -1.63 | ✓ | ✗ | 8.52 | ✓ |
|  | d2 | 0.40 | -1.17 | ✗ | ✓ |  |  |
|  | d3 | 0.71 | -0.86 | ✗ | ✓ |  |  |
| V$_{MoS3}$ | d1 | -0.25 | -1.78 | ✓ | ✗ | 11.81 | ✓ |
|  | d2 | 0.89 | -0.64 | ✗ | ✓ |  |  |
|  | d3 | 0.99 | -0.53 | ✗ | ✓ |  |  |
| V$_{MoS6}$ | type 1 | < 0.50 | > 1.50 | ✓ | ✗ | 21.41 | ✗ |
|  | type 2 | < 1.00 | > 1.00 | ✗ | ✗ |  |  |
|  | type 3 | > 1.75 | < 0.25 | ✗ | ✓ |  |  |
| O$_2$ at interface | d1 | -0.99 | -2.45 | ✓ | ✗ | 0.68 | ✓ |
|  | d2 | -0.55 | -2.00 | ✓ | ✗ |  |  |
|  | d3 | -0.85 | -2.31 | ✓ | ✗ |  |  |
| H$_2$O at interface |  | -3.42 | -4.91 | ✗ | ✗ | 0.61 | ✗ |
| O$_2$ in MoS$_2$ |  | -0.37 | -2.01 | ✓ | ✗ | 2.35 | ✓ |
| O$_2$ at surface |  | 1.11 | -0.41 | ✗ | ✓ | 2.25 | ✓ |

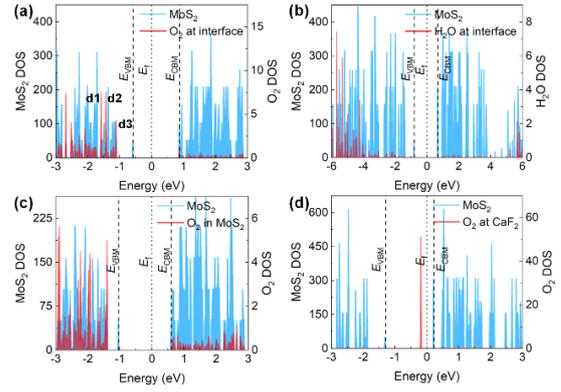

**Fig. 4** The energy level distribution of different adsorbents. The contributions of defect energy levels in Figures (a) to (d) correspond to the cases (a) to (d) in Fig. 4, respectively.

It has been mentioned in previous reports that the hysteresis of CaF$_2$-MoS$_2$ devices can be reduced after they are heated and dried.[13] This indicate that molecules had been adsorbed during device preparation, so the activity of these adsorbates need to be discussed. Fig. 4(a) shows the adsorption of O$_2$ in the CaF$_2$-MoS$_2$ interface, and three defect levels that denoted by d1, d2 and d3 are observed. They are only 1 eV, 0.85 eV and 0.54 eV below VBM, respectively. Therefore, they will be active hole traps in p-MOSFETs. In contrast, the adsorption of water molecules on the interface is much less important. It can be seen from Fig. 4(b) that there is no obvious defect state near the band edge of MoS$_2$. To further check the importance of oxygen, we studied the oxygens that adsorbed in other positions. Fig. 4(c) shows the situation that the oxygen molecules are adsorbed in the interlayer of MoS$_2$. It can be seen that the defect state is only 0.37 eV below VBM, which will trap holes easily and thus affects the device performance. Fig. 4(d) shows the case that the oxygen is adsorbed on the surface of CaF$_2$. An occupied defect state that is close to CBM rather than CBM is seen. Considering that the negative gate voltage in a p-FET will drag the defect level down towards the VBM, the oxygen on the CaF$_2$ surface will be very active hole trapping centers with large gate voltage.

To exhibit the importance of different defects more clearly, Table 1 summarizes and compares the information of all defects. The defect levels that are more than 1 eV away from the MoS$_2$ band edge are regarded as electronically unimportant.[41–43] Moreover, the formation energy/adsorption energy is considered to provide an overall evaluation of their importance.

Now we study the MoSi$_2$N$_4$/CaF$_2$ system. MoSi$_2$N$_4$ is a 2D material with 7 atomic layers. One Mo atomic layer lies in the middle while two Si-N-Si tri-layers lie on top and bottom surface symmetrically. Vacancy defects caused by the shedding of N (Fig. 5a) and Si (Fig. 5b) atoms on the surface layer are the primary problems to be considered. At the same time, the influence of oxygen molecules (Fig. 5c) and water molecules (Fig. 5d) adsorption during device manufacturing is also considered here. The atoms highlighted in red in the figure represent defects and adsorption sites.

For the N vacancy (V$_N$) (Fig. 6a), two defect levels are induced into the band gap, of which the half-occupied d1 state is 0.98 eV above VBM and the empty d2 state is 0.45 eV below CBM. Such small energy barriers make them very active hole/electron trapping centers. In contrast, the V$_{Si}$ defect induce no defect levels close to the CBM, as is shown in Fig. 6(b), but it induces many defect levels below the VBM. Specially, the electrons in VBM have spontaneously transferred to the defect sates, making the Fermi level shifted below the VBM, and making the CaF$_2$-MoSi$_2$N$_4$ as a whole P-type heterostructure. Interestingly, the adsorption of an oxygen in the CaF$_2$-MoSi$_2$N$_4$ interface has very similar effect, as is shown in Fig. 6(c), the electrons in VBM are spontaneously captured by the oxygen, and the MoSi$_2$N$_4$ becomes a p-type material. If the oxygen density is high, this will greatly impair the performance and reliability of the device. In comparison, the adsorption of water molecules in the interface does not have such effect, as is shown in Fig. 6(d). The water related defect energy level is far away from the band edge of MoSi$_2$N$_4$. This further confirms that the water molecules adsorption is less important than oxygen adsorption in impacting device performance and reliability.

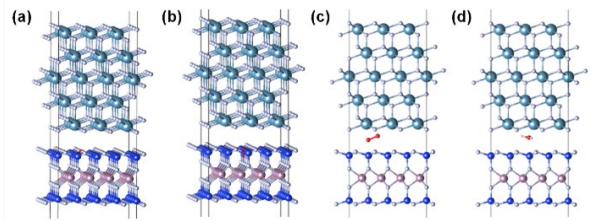

**Fig. 5** Common defects and adsorption in CaF$_2$-MoSi$_2$N$_4$. (a) and (b) are V$_N$ and V$_{Si}$ in MoSi$_2$N$_4$ respectively. Figures (c) and (d) are the adsorption of water molecules and oxygen molecules.

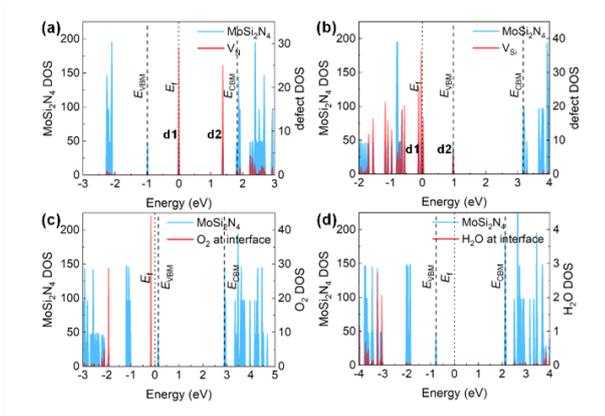

**Fig. 6** The energy level distribution of common defects and adsorption in CaF$_2$-MoSi$_2$N$_4$. The contributions of defect energy levels in Fig. (a) to (d) correspond to the cases (a) to (d) in Fig. 4, respectively.

To present the importance of different defects more intuitively, Table 2 summarizes and compares the information of all defects in CaF$_2$-MoSi$_2$N$_4$ system.

## 4. CONCLUSION

In conclusion, we have investigated the various defects and adsorbates in CaF$_2$-based 2D material MOSFET structures to distinguish their importance in degrading the device performance and reliability. First, the intrinsic defects in channel materials, including the V$_S$ and V$_{Mo}$ in MoS$_2$, and V$_{Si}$ and V$_N$ in MoSi$_2$N$_4$ are very active charge trapping centers. Second, the adsorbed oxygen molecule in the channel/CaF$_2$ interface or CaF$_2$ surface is very important trap centers, and they can even spontaneously change the MoSi$_2$N4 to p-type. Third, the adsorbed water molecules are very inactive in capture charges, and thus is much less important in affecting device performance. An elaborate Table that comparing the detailed properties of different defects is provided so that both the researchers in experiment and in theory can refer to easily. These results mean that the exclusion of adsorbates in device fabrication is as important as growing high-quality channel material to obtain better device performance.

**Tab. 2** Importance of different capture centers in CaF$_2$-MoSi$_2$N$_4$.

| Defect Types | Defect State | ΔE-VBM (eV) | ΔE-CBM (eV) | NBTI Important | PBTI Important | Formation/ Adsorption energy (eV) | Comprehensive Importance |
|---|---|---|---|---|---|---|---|
| V$_N$ | d1 | 0.98 | -1.83 | ✓ | ✗ | 5.97 | ✓ |
|  | d2 | 2.36 | -0.45 | ✗ | ✓ |  |  |
| V$_{Si}$ | d1 | -1.01 | -3.23 | ✗ | ✗ | 11.15 | ✓ |
|  | d2 | 0.00 | -2.22 | ✓ | ✗ |  |  |
| O$_2$ at interface |  | -0.32 | -3.07 | ✓ | ✗ | 0.19 | ✓ |
| H$_2$O at interface |  | -2.29 | -5.17 | ✗ | ✗ | 0.34 | ✗ |

## Author Contributions

Zhe Zhao: Conceptualization, Methodology, Data collection, Writing – original draft. Tao Xiong: Writing – review & editing. Jian Gong and Yue-Yang Liu: Supervision, Writing – review & editing.

## Conflicts of interest

There are no conflicts to declare.

## Acknowledgements

This work was financially supported by the National Natural Science Foundation of China Grant No. 12004375, in part by the National Natural Science Foundation of China Grant No. 62174155, the National Natural Science Foundation of China Grant No. 62004193, the National Natural Science Foundation of China Grant No. 62125404, the Inner Mongolia Natural Science Foundation No. 2023ZD27, and the National Natural Science Foundation of China Grant No. 11964022.